\documentstyle[preprint,eqsecnum,aps]{revtex}

\begin{document}
\title{Vacuum Polarization Contribution to the Astrophysical 
S-Factor}
\author{A.B. Balantekin$^a$, C. A. Bertulani$^b$ and M.S. 
Hussein$^c$\\}
\address{
$^a$ University of Wisconsin, Department of Physics\\
Madison, WI 53706, USA. E-mail: baha@nucth.physics.wisc.edu\\ 
$^b$ Instituto de F\'\i sica, Universidade Federal 
do Rio de Janeiro \\
21945-970 \ Rio de Janeiro, RJ, Brazil. E-mail: 
bertu@if.ufrj.br\\ 
$^c$ Universidade de S\~ao Paulo, Instituto de F\'\i sica\\
05389-970 \ S\~ao Paulo, SP, Brazil. E-mail: hussein@if.usp.br}
\maketitle

\begin{abstract}
We study the effect of vacuum polarization in nuclear reactions 
of
astrophysical interest. This effect has the opposite sign 
compared to the
screening by the atomic electrons. It is shown that vacuum 
polarization
further increases the longstanding differences between the 
experimental data
of astrophysical nuclear reactions at very low energies and the 
theoretical
calculations which aim to include electron screening, albeit by a 
small 
amount.
\end{abstract}


Understanding the dynamics of fusion reactions at very low 
energies is
essential to understand the nature of stellar nucleosynthesis. 
These
reactions are measured at laboratory energies and are then 
extrapolated to
thermal energies. This extrapolation is usually done by 
introducing the
astrophysical S-factor: 
\begin{equation}
\sigma \;(E)={\frac 1E}\;S(E)\;\exp (-2\pi \eta )\ ,  
\label{sig1}
\end{equation}
where the Sommerfeld parameter, $\eta ,$ is given by$\;\eta 
=Z_1Z_2e^2/\hbar
v.$ Here $Z_{1,}$ $Z_2,$ and $v$ are the electric charges and the 
relative
velocity of the target and projectile combination.

The term $\exp (-2\pi \eta )\ $is introduced to separate the 
exponential
fall-off of the cross-section due to the Coulomb interaction from 
the
contributions of the nuclear force. The latter is represented by 
the
astrophysical S-factor which is expected to have a very weak 
energy
dependence. The form given in Eq. (1) assumes that the electric 
charges on
nuclei are ``bare''. However, neither at very low laboratory 
energies, nor
in stellar environments this is the case. In stars the bare 
Coulomb
interaction between the nuclei is screened by the electrons in 
the plasma
surrounding them. A simple analytic treatment of plasma screening 
was
originally given by Salpeter\ \cite{Sal54}. In most cases of 
astrophysical
interest Salpeter's treatment still remains to be a sufficient 
approximation 
\cite{Bah96}. In the very low energy laboratory experiments the 
bound
electrons in the projectile or the target may also screen the 
Coulomb
potential as the outer turning point gets very large (%
\mbox{$>$}
500 fm). As experimental techniques improve one can measure the 
cross
section in increasingly lower energies where the screened Coulomb 
potential
can be significantly less than the bare one. This deviation from 
the bare
Coulomb potential should manifest itself as an increase in the 
astrophysical
S-factor extracted at the lowest energies. This enhancement was 
indeed
experimentally observed for a large number of systems \cite
{Eng88,Eng92,Ang93,Pre94,Gre95}. The screening effects of the 
atomic
electrons can be calculated \cite{Ass87} in the adiabatic 
approximation at
the lowest energies and in the sudden approximation at higher 
energies with
a smooth transition in between \cite{Sho93}. Contributions from 
the nuclear
recoil caused by the atomic electrons are expected to further 
increase the
screening effect for asymmetric systems \cite{Sho93,Bri81} . In 
almost all
the cases observed screening effects are found to be equal to or 
more than
the theoretical predictions. Recently including improved energy 
loss data
for atomic targets is shown to lead agreement between theory and 
data \cite
{newkarl}, however the situation is still not resolved for 
molecular and
solid targets. Electron screening enhancement was not observed 
for the
heavier symmetric system $^3He(^3He,2p)^4He$ \cite{Kra87} which 
is expected
to have about 20\% enhancement at the energies studied. Recent 
measurements 
\cite{rol} have not yet clarified the effects of electron 
screening in this 
reaction. A mechanism which
reduces the screening enhancement for this system (and possibly 
for other
systems with large values of $Z_1Z_2$ and the reduced mass) seems 
to be
needed. In this note we show that the contributions from the 
polarization of
the vacuum alone cannot achieve this task. The effects of the 
polarizability
was previously investigated in Ref. \cite{Bau77} 
for elastic scattering below the Coulomb barrier and in Ref. 
\cite{son} 
for subbarrier fusion reactions using the formalism developed by 
Uehling \cite{Ueh35}. Effects of vacuum polarization in 
$^{12}C-^{12}C\;$scattering at 4 MeV was
subsequently experimentally observed \cite{Vet89}.

Vacuum polarization {\bf increases} the electromagnetic potential 
between
two like charges. Like the Coulomb potential itself, the increase 
due to
vacuum polarization is also proportional to the product of the 
charges \cite
{Ueh35}. Vacuum polarization contribution increases almost 
exponentially as
the two charges get closer. The Coulomb interaction is smaller 
for
asymmetric systems than for symmetric systems of comparable size. 
On the
other hand, the nuclear force tends to extend farther out for 
asymmetric
systems because of the extra neutrons. Consequently for 
asymmetric systems
the very tail of the nuclear force can turn the relatively weak 
Coulomb
potential around to form a barrier at a considerable distance 
from the
nuclear touching radius. For symmetric systems, however, the 
location of the
barrier is further inside where the vacuum polarization 
contribution is
stronger. We show that the resulting increase in vacuum 
polarization is
nevertheless not sufficiently large to make an appreciable 
contribution to
the extracted astrophysical S-factor. For light symmetric systems 
with small
values of $Z_1Z_2$ this effect should be negligible. Indeed, for 
the $pp$
reaction the vacuum polarization contribution was shown to be 
very small 
\cite{Kam94}. Similarly the measured S-factor for the $d(d,p)^3H$ 
reaction 
\cite{Gre95} agrees well with theoretical calculations of atomic 
screening 
\cite{Bra90}. On the other hand one may expect that already for 
the $%
^3He(^3He,2p)^4He$ reaction the increase in the potential due to 
the vacuum
polarization could be large enough to counter the decrease due to 
electron
screening. We show that this is not the case.

The vacuum polarization potential is according to Uehling 
\cite{Ueh35} given
by

\[
V_{Pol}(r)=\frac{Z_1Z_2e^2}r\;\frac{2\alpha }{3\pi }\;I%
{2r \overwithdelims() \lambda _e}
\;, 
\]
where $\alpha =1/137$ is the fine structure constant, and 
$\lambda _e=386$
fm is the Compton wavelength of the electron. The function $I(x)$ 
is given by

\[
I(x)=\int_1^\infty e^{-xt}\;(1+\frac 
1{2t^2})\;\frac{\sqrt{t^2-1}}{t^2}%
\;dt\;. 
\]
As shown by Pauli and Rose \cite{Pau36} this integral can be 
rewritten as

\[
I(x)=\alpha (x)K_0(x)+\beta (x)K_1(x)+\gamma (x)\int_x^\infty 
K_0(t)\;dt\;, 
\]
where

\begin{eqnarray*}
\alpha (x) &=&1+\frac 
1{12}x^2\;,\;\;\;\;\;\;\;\;\;\;\;\;\;\;\beta (x)=-%
\frac 56x(1+\frac 1{10}x^2)\;, \\
\gamma (x) &=&\frac 34x(1+\frac 
19x^2)\;,\;\;\;\;\;\;\;\;\;\;\;\text{with}%
\;\;x=2r/\lambda_e \;.
\end{eqnarray*}
In Ref. \cite{Bau77} it was shown that the modified Bessel 
functions $K_0$
and $K_1$ as well the integral over $K_0$ can be expanded in a 
very useful
series in Chebyshev polynomials which converge rapidly and for 
practical
purposes only a few terms $(\approx 5-10)$ is needed, allowing a 
very fast
and accurate computation of the Uehling potential.

In Figure 1 we plot the Coulomb potential and the vacuum 
polarization
potential for $Z_1Z_2=1$. Both the Coulomb potential and the 
screening
potential scale with the product $Z_1Z_2$. However, the vacuum 
polarization
potential has a stronger dependence on the nuclear separation 
distance.

To calculate the effects of screening by the atomic electrons, 
and the
effect of vacuum polarization we use for simplicity the (s-wave) WKB 
penetrability
factor

\begin{equation}
P_{C+Scr+Vac.Pol.}(E)={\exp \{-}\frac 2\hbar \int_{R_n}^{R_C}dr\ 
\left[
2m\left( V_c+V_{pol}-E^{\prime }\right) \right] ^{1/2}\}\ ,  
\label{tmat}
\end{equation}
where $E^{\prime }=E+U_e,$ with $E$ being the relative energy 
between the
nuclei, and the atomic screening correction, assumed to be a 
constant, is
given by $U_e$. The limits of the integral are the nuclear 
radius, $R_n,$
where the nuclear fusion reaction occurs, and the classical 
turning point in
the Coulomb potential, $R_C=Z_1Z_2e^2/E^{\prime }.$ At very low 
energies the
inferior limit $R_n$ is not important when vacuum polarization is 
neglected
(the exponential factor in Eq. 1 can be obtained with 
$V_{pol}=0,$\ and $%
R_n\rightarrow 0,$ in Eq. 2). However, since the vacuum 
polarization
potential has a strong dependence on the nuclear separation 
distance, being
much stronger at shorter distances, its effect is very much 
dependent on the
choice of this parameter. For the sake of simplicity, we use the
conventional approximation $R_n=1.2\;(A_1^{1/3}+A_2^{1/3})\;$fm 
for the
nuclear radius. In Table 1 we show the ratio between the 
penetrability
factor through the Coulomb barrier and the penetrability factor 
including
atomic screening, $P_{C+Scr}(E)/P_C(E)$, and also including the 
effect of
vacuum polarization, $P_{C+Scr+Vac.Pol.}(E)/P_C(E)$. The energy 
$E$ chosen
is the lowest experimental energy for each reaction. The atomic 
screening
corrections $U_e$ were calculated in the adiabatic approximation, 
given by
the differences in electron binding energies between the 
separated atoms and
the compound atom \cite{Sho93}. We see that the effect of vacuum
polarization is small, but non-negligible for some reactions. 
Moreover, it
increases the discrepancy between the value of the screening 
potential
required to explain the experimental data and the theoretical 
calculations
of this potential as illustrated in Table 1.

In conclusion, we showed that the vacuum polarization 
contribution to the
astrophysical S-factor never exceeds a few percent, but may be 
significant
in extrapolating the measured S-factor to lower energies. 
Although its
contribution is not comparable to that of sub-threshold 
resonances and
electron screening, vacuum polarization is one of the many 
factors that may
contribute to the weak energy dependence of the S-factor. Vacuum
polarization effects are sensitive to the inner turning point of 
the
potential barrier, hence to the diffuseness of the nuclear 
potential
employed.

\bigskip\bigskip
\noindent{\bf Acknowledgments}

\medskip

We thank K. Langanke for useful comments. 
This work was supported in part by MCT/FINEP/CNPQ(PRONEX) under 
contract No.
41.96.0886.00, in part by the U.S. National Science Foundation 
Grant No.
PHY-9605140, in part by the FAPESP under contract number 96/1381-0,
and in part by the University of Wisconsin Research 
Committee
with funds granted by the Wisconsin Alumni Research Foundation.

\newpage
\bigskip

\begin{tabular}{|l|l|l|l|l|}
\hline
Reaction & $E_{\min }[keV]$ & $U_e[eV]$ & $P_{C+Scr}/P_C$ & $%
P_{C+Scr+Vac.Pol.}/P_C$ \\ \hline
$D(d,p)T$ & 1.62 & 20 & 1.163 & 1.150 \\ \hline
$^3He(d,p)^4He$ & 5.88 & 119 & 1.331 & 1.303 \\ \hline
$D(^3He,p)^4He$ & 5.38 & 65 & 1.195 & 1.173 \\ \hline
$^3He(^3He,2p)^4He$ & 25 & 292 & 1.196 & 1.159 \\ \hline
$^6Li(p,\alpha )^3He$ & 10.74 & 186 & 1.258 & 1.231 \\ \hline
$^7Li(p,\alpha )^4He$ & 12.70 & 186 & 1.197 & 1.173 \\ \hline
$^6Li(d,\alpha )^4He$ & 14.31 & 186 & 1.218 & 1.186 \\ \hline
$H(^6Li,\alpha )^3He$ & 10.94 & 186 & 1.250 & 1.224 \\ \hline
$H(^7Li,\alpha )^4He$ & 12.97 & 186 & 1.191 & 1.167 \\ \hline
$D(^6Li,\alpha )^4He$ & 15.89 & 186 & 1.183 & 1.153 \\ \hline
$^{10}B(p,\alpha )^7Be$ & 18.70 & 346 & 1.376 & 1.338 \\ \hline
$^{11}B(p,\alpha )^8Be$ & 16.70 & 346 & 1.461 & 1.419 \\ \hline
\end{tabular}

{\bf Table Caption:}

Lowest experimental energies, $E_{\min }$, energy corrections 
\cite{Bra90}
due to the screening by the atomic electrons, $U_e$, and 
enhancement factors
for the nuclear reaction: (a) due to atomic screening, 
$P_{C+Scr}/P_C$, and
(b) due to the combined effect of atomic screening and vacuum 
polarization, $%
P_{C+Scr+Vac.Pol.}/P_C.$

\bigskip 
{\bf Figure Caption}\\

{\bf Fig. 1} - Comparison between the Coulomb potential and the 
vacuum
polarization potential as a function of the nuclear separation 
distance for $%
Z_1Z_2=1.$ The vacuum polarization potential has been multiplied 
by a factor
1000 in order to be visible in the same plot.

\end{document}